\documentclass[a4paper,11pt,conference]{IEEEtran}
\usepackage{comment}

\usepackage{graphicx}
\usepackage{placeins}
\usepackage{float}
\usepackage{subfigure}
\usepackage{xcolor}
\ifCLASSINFOpdf
\else
\fi
\begin{document}

%\title{Non-Consistent and Consistent Heteroskedasticity as Respective Signatures of Association and "Probable Causality" for Age-Related Genes}

\title{Heteroskedasticity as a Signature of Association for Age-Related Genes}

% \begin{comment}

\author{\IEEEauthorblockN{Salman Mohamadi}
\IEEEauthorblockA{Computer Science \& Electrical Engineering\\
West Virginia University,
Morgantown, WV\\
sm0224@mix.wvu.edu}
\and
\IEEEauthorblockN{Donald A. Adjeroh}
\IEEEauthorblockA{Computer Science \& Electrical Engineering\\
West Virginia University,
Morgantown, WV\\
donald.adjeroh@mail.wvu.edu}}

% \end{comment}
% \and
% \IEEEauthorblockN{James Kirk\\ and Montgomery Scott}
% \IEEEauthorblockA{Starfleet Academy\\
% San Francisco, California 96678--2391\\
% Telephone: (800) 555--1212\\
% Fax: (888) 555--1212}
% }


%Telephone: (800) 555--1212, Fax: (888) 555--1212}


\maketitle

%\vspace{-10em}
\begin{abstract}
Human aging is a process controlled by both genetics and environment. Many studies have been conducted to identify a subset of genes related to  aging from the human genome. Biologists implicitly categorize age-related genes into genes that cause aging and genes that are influenced by aging, which resulted in both causal inference and inference of associations studies. While inference of association is better explored,  causal inference and computational causal inference, remains less explored.
In this work, we are primarily motivated to tackle the problem of identifying genes associated with aging, while having a brief look into genes with probable causal 
relations, both from a computational perspective. 
Specifically, we form a set of hypotheses and accordingly, 
introduce a data-tailored framework for inference. First we perform linear modeling on the expression values of age-related genes, and then examine the presence of heteroskedastic properties in the residual of the model. We evaluate this framework and our results suggest that, 1) presence of heteroskedasticity in these residuals is a potential signature of association for age-related genes, and 
2) consistent heteroskedasticity along the human life span could imply some sort of causality. 
To our knowledge, along with identifying age-associated genes, this is the first work to propose a framework for computational causal inference on age-related genes, using a dataset of human dermal fibroblast gene expression data. Hence the results of our simple, yet effective approach can be used not only to assess future age-related genes, but also as a possible criterion to select new associative or potential causal genes with respect to aging. 
\end{abstract}

\section{Introduction}
\label{sec:intro}
The process of human aging has gained a huge attention over the years due to its implications on many other healthcare domains such as public health,  
% health policy-making, 
health policy, and human longevity. Simply put, aging is a complex physiological process which comes with progressive physiological decline and consequently, more risk of mortality \cite{b1}. Prior work on transcriptome of human dermal fibroblast strongly suggest the relationship between variations in gene expression levels and the aging process \cite{b2,b3, b4}. These variations in gene expression level could either come from the genetics or the environment. Accordingly, one can make distinctions between the types of age-related genes. That is, age-related genes do not all function in the same way -- \textbf{some cause (influence) aging, while others are influenced by the aging process.}

In fact, several questions in healthcare research may call for an 
%are encouraged by 
investigation of both association and causality 
%rather than association, 
because certain practical solutions often 
lie with the cause of the effect \cite{b5,b6}. Therefore, as a key takeaway, in some application scenarios, %it is very 
(be it in computational biology research or in medical research), it may be helpful to discriminate the type of relationship.
Though there is significant work from the wet-biology laboratory experiments, there is very limited computational work on this causal inference  problem.    
%Unlike the vet lab experiments, diving into sparse literature of computational work on genomic data to identify age-related genes, informed us of lack of attention to inference of association. 
%However, hoping to encourage computational causal inference in parallel with vet lab experiment on causality, .
In this paper, we develop one of the first known computational frameworks (to our knowledge) on causal inference 
%that investigates the causal relationship 
between age-associated genes and the aging process using human dermal fibroblast gene expression data. 
% \vspace{-1.2em}

\subsection{Gene Expression and Human Aging}
% \vspace{-.7em}

Human DNA contains close to 30,000 genes which conserve instructions to initiate and control different functions.
%with the goal of attaining functional product. 
The process of converting the information of a given gene to a functional product such as proteins, namely, gene expression, was identified several decades ago and can be quantitatively measured with high accuracy \cite{b7}. From a general point of view, an organism can be characterized based on the levels of expression of certain  genes, under specified experimental conditions. Aside from human gene expression data, there is a rich literature on analysis of gene expression data from animals \cite{b47,b48,b49}. Thus, gene expression is viewed as an important part in converting information stored in DNA to observable traits in an organism. Hence, computational analysis or pattern recognition on gene expression data could enhance our understanding of functional mechanisms in an organism \cite{b8}. Mechanisms behind aging also could be studied within the same context \cite{b4}, e.g., to better understand the cause or effect of aging using the expression level of a certain subset of relevant genes. Variations in gene expression along the human lifespan could have implications on the possible functionality of a given gene. In this work, we analyse such variations using a dataset of dermal fibroblast gene expression data.
% \makeatletter\typeout{\Gin@extensions}\makeatother
\begin{figure*}[t]
\centering{
\includegraphics[scale=01,width=3.5in]{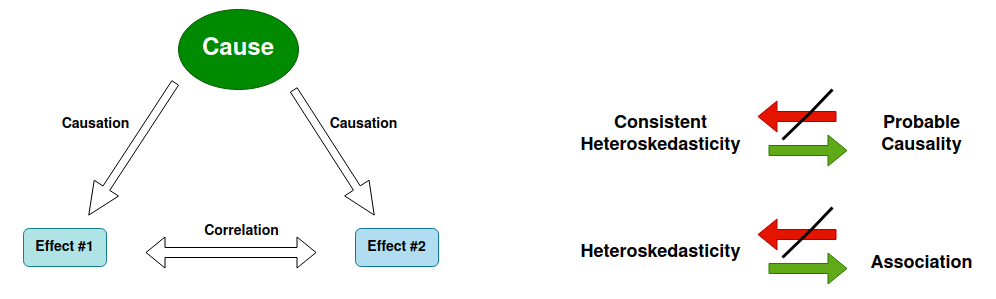}
% \includegraphics[scale=1]{Fig1}
% 		\vspace{-2.em}

		\caption{Key concepts and key hypotheses. Left: causality vs correlation. Causal inference is related but different from  inference of association, adapted from \cite{b9}. 
		Right:  Set of hypotheses. The model is only capable of performing inference on genes with variation in their expression values over time, as opposed to those with no variation.
		% \textcolor{red}{For this figure, everyone knows about the explosive growth of COVID-19 cases or deaths. For this paper, it will be better to show the growth of available COVID-19 sequences, and or the growth of COVID-19 variants.
		\label{Fig.0}}}
% 		\vspace{-1.75em}
\end{figure*}
% \vspace{-.5em}
% \vspace{-1.2em}
% \begin{figure}[!t]
% \centering
% \includegraphics[width=2.5in]{CausalvsAssociative.eps}
% where an .eps filename suffix will be assumed under latex,
% and a .pdf suffix will be assumed for pdflatex; or what has been declared
% via \DeclareGraphicsExtensions.
% \caption{Simulation results for the network.}
% \label{fig_sim}
% \end{figure}

\subsection{Causal Inference and Inference of Association}
% \vspace{-.7em}

Generally speaking, causality is a complex philosophical concept. From a statistical point of view, it is concerned with whether a variable is or is likely to be the cause of another variable, which consequently enables a better prediction of the affected variable \cite{b5}. Causation is different from correlation as shown in Fig. \ref{Fig.0}. In fact, causal reasoning is often difficult to perform in practice, and is also challenging to formulate in mathematical context. Therefore, one should be seriously concerned about mis-attribution in the context of correlation, independent effect, and causality. One of the pioneer methods which represents the concept of causal relationship in the context of time series (such as our data), is the Granger causality test. This is a statistical hypothesis testing approach, where causality is interpreted as the ability of one variable to predict the future of another variable better than the variable itself \cite{b10}. 
%An important aspect of investigations in 
A key aspect in studying causal relationships is \textbf{model specification}.
%, as the goal is to aim at causal relationship and not other types of relationships. 
In this work, we employ a model that enables us to focus on causal inference along with inference of association; drawing a line between causal and statistical concepts in the sense that causality holds causal assumptions that, unlike the statistical concepts, may require knowledge beyond the joint distributions \cite{b9}.

% In his book, Judea Pearl \cite{b9} tried to reconcile previous methods to causal inference and develop a unified mathematical formulation. However, the work draws a line between causal and statistical concepts in the sense that causality holds causal assumptions that, unlike the statistical concepts, may require knowledge beyond the joint distributions. 
% \vspace{-1.5em}
\subsection{Motivation}
% \vspace{-.7em}
Statistical concepts, such as correlation, dependence, and conditional independence 
%which are all representable 
can be captured by the joint probability distribution. Causal reasoning, on the other hand, may require further guidance using prior knowledge of domain experts. Biology literature on aging process and genomic data confirm complex variations in expression level of some of the age-associated genes, which could differ even among people with the same age and similar background \cite{b1,b11}. The behaviour of variance as a second moment reflects the complexity and can be used to analyze the nature of complexity. This motivates us to further explore the behaviour of this disturbance in terms of heteroskedasticity as a discriminative characteristic. 
% We hypothesize that this could be a discriminative measurement for causal reasoning. 
% Moreover, some studies suggested investigation of heteroskedasticity as a very good prediction as well as characterization of cause of phenomenon. 

%Along with all mentioned above, it is important to note that approaching gene expression data with causal inference rather than inference of association, could also be helpful to re-verification on the subset of known age-related genes to filter out possible outliers. 
As Engle (inventor of ARCH) and Granger (inventor of Granger causality) emphasized in their paper on cointegrating vector approach \cite{b12}, non-stationary time series data might show spurious correlation since standard detrending techniques in linear regression models may result in data that are still non-stationary.
Thus, using causal inference (rather than just inference of association)  
%approaching gene expression data analysis from the viewpoint of causal inference rather than inference of association, 
in gene expression analysis 
could also be helpful in computational validation of a subset of known age-related genes, and to filter out possible outliers. 
Our view of causality in this work is primarily from this perspective of Granger causality, considering the gene expression data as a time series.    

Relevant examples of studies on age-related genes using computational methods include work by Srivastava et al \cite{b13}, Uddin et al \cite{b14},  Avelar et al \cite{b15}, and Mohamadi et al \cite{b15b1,b15b2}. These mostly applied machine learning or signal processing algorithms to different datasets. Causal inference have also been used in other genome-wide studies, for example   \cite{b16,b17,b18}. However, these were not on causal genes with respect to aging. 

Our work is also related to 
%We also mention It is also worth mentioning that we have studied a line 
recent studies on causal inference in machine learning,  such as \cite{b26,b27,b28, b29,b30,b46}. These have focused mainly on developing \textbf{probabilistic models} to be trained for the inference and in part, benefit from availability of large datasets, or even redundant instances of the same sample.
%subject. 
However, referring to  \cite{b46} which clearly investigated the aspect of the challenge, and also due to the very limited available data in our problem setting, these machine learning based approaches are not easily applicable to our problem of identifying causal genes related to aging. Thus, we take a different approach to causal inference on our data, following primarily statistical approaches used in econometrics.  
%For this reason, we find it necessary to come up with a fairly different probabilistic modeling for casual inference on the data.

%however, mentioned examples on aging and many other work are performed from the perspective of inference of association rather than causal inference. 

% \textcolor{Red}{Example of prior work on age-related genes and causal genes. Three things:

% -- age-related genes and associations

% -- causal inference in general, and in genes/genetics

% -- establish that no one has ever used heteroskedasticity for causal inference on age-related genes. 
% }

In this work, first we apply a well-known linear modeling approach, namely auto-regressive integrated moving average (ARIMA), on the vectors of gene expression data individually. Then, we perform heteroskedasticity test on the residuals of the ARIMA model. We develop our hypothesis and study possible connections between our initial results and literature on related prior biological lab experiments. Then we evaluate the hypothesis, and discuss how to extend the work to the broader problem of finding novel age-associated genes, not just causal genes. The main contributions of the paper are as follows:
\begin{itemize}
    \item We develop the first causal inference framework for identifying causal age-related genes, using gene expression data from human dermal fibroblast;
%    \item After discerning the drawback of the general line of machine learning-based causal inference with regard to our dataset, we formalized hypothesis based on deep filed knowledge of the type of data. Then we came up with a specialized framework for testing the hypothesis on this type of data.
    \item We extend our primarily causal inference framework to inference of association, which allows for further exploration on finding genes associated with aging; 
    \item{We show results on a gene expression dataset that demonstrates  the performance of the proposed framework. Then, we provide some form of external support of our computational results using evidence from the literature on related wet-biology experiments. }
    \end{itemize}

% \vspace{-1.2em}
\section{Problem Statement}
\label{sec:format}
% \vspace{-.7em}
%In this section first, 
We introduce the datasets used and briefly define the problem and the assumptions in our inference, such as, independence. We also emphasize how the complexity of the aging process makes the data distinct from other non-biological data and what would be considered as good results, hence support the significance of our results.
% \vspace{-1.6em}
\subsection{Datasets}
% \vspace{-.7em}
We use three datasets in this work.
First, we use a dataset of gene expression values from human dermal fibroblast transcriptome 
%which is developed and 
reported in \cite{b4}, and also used in \cite{b15b1,b15b2}. 
The  database consists of a matrix of gene expression values from 27,142 genes across 143 individuals. Each row represents the expression values from one gene across 143 subjects in order of age (from age $<$1 year to age 94). Each column maintains the expression values for each of  27,142 different genes for a given individual. 
Ideally, data collection would include the expression values for any given gene across the lifetime of each individual. However, this is not practical, and the measurements for expression values of each gene in the data set are attained from 143 different individuals with age ranging from less than 1 to 94 years old. Previous studies suggest that only about ${2\%}$ of these genes are supposed to be age-related \cite{b1}. Hence, most  genes may not be related to aging, and the main part of our computational experiments revolves around the small subset of genes that potentially relates to aging.
Second, using a well-known dataset collected based on the literature\footnote{https://genomics.senescence.info/genes/index.html}, we have a list of 69 genes identified as presumably having a causal relationship with aging. 
Third, from the same website, we have a longer list of 550 other age-related genes, of which 275 are indicated to be likely influenced by aging, and the other are only indicated to be associated with aging, with no information on  causality \cite{b11,b1}.
% \vspace{-1.6em}
\subsection{Inference}
% \vspace{-.7em}
We are interested in recognizing both causal and non-causal age-related genes. That is, genes that 
%have causal relationship with aging in terms of some degree of 
contribute to aging or are affected by aging. Previous studies mostly aimed for identifying age-associated genes \cite{b13,b14,b15, b15b1}, and not necessarily on the direction of the association. Hence, there is  sparse literature on identifying \textbf{genes responsible for aging}. Even the limited literature is mostly concerned with wet lab experiments \cite{b11}, rather than computational inference from the data. This might, in part, be explained by the lack of available data for this type of analysis. 

Here, the dataset consists of 27,142 genes. However, apart from the 69 age-related genes that are indicated to be consistently over-expressed or under-expressed with age \cite{b11}, and contributing to aging, a subset of 550 other genes are identified as age-related. 
%genes which will be considered for causal inference. 
We apply our approach on the 69 causal genes to verify its capability, and then on the 550 other age-related genes, and finally on the remaining 26,523  (i.e., 27142-550-69) genes to identify possibly novel age-related genes. 

Note that similar to many other physiological mechanisms, aging is a complex mechanism. It is expected that the subset of genes that drive aging may function interdependently due to their possible  interactions. To capture this potential network effect, causality needs to be investigated jointly. However, to simplify the problem for practical consideration, 
%without undermining the backbone of our causal inference, 
for this initial investigation, we assume that genes function independently. %Accordingly we perform the causal inference on each gene independent of the others.

\begin{figure*}[t]
	\centering{
		\includegraphics[scale=.45]{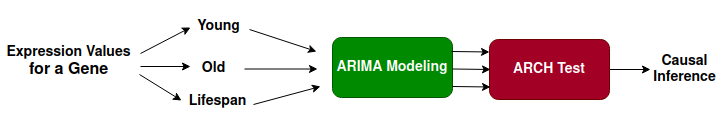}
		%\includegraphics[scale=1]{Fig1}
% 		\vspace{-1.em}

		\caption{Pipeline of the framework for a given gene. 
		% \textcolor{red}{For this figure, everyone knows about the explosive growth of COVID-19 cases or deaths. For this paper, it will be better to show the growth of available COVID-19 sequences, and or the growth of COVID-19 variants.
		\label{Fig.001}}}
% \vspace{-1.5em}
\end{figure*}
% \vspace{-1.6em}

\section{Methodology}
% \vspace{-.7em}
\label{sec:pagestyle}

%\subsection{Theoretical Foundation of The Hypothesis}
\subsection{Devising an approach}
 We motivate our approach by investigating possible alternatives such as machine learning-based casual and associative inference. Recent work such as \cite{b46} has %amazingly
 shed light on the possible challenges with machine learning tools for 
 %biological causal inference 
 causal inference on biological data. There are two main reasons why traditional machine learning approaches generally do not fit our problem. First, to perform well these approaches generally need very large datasets with appropriate labels, whereas this type of data is not currently available for our problem. Second, in general, the dataset needs to be unbiased, meaning that it should have almost the same number of different classes (here causal, associative, and age-unrelated gene classes). However, we 
 %already 
 also know that this is not possible as the literature suggests that only around $2\%$ of genes are expected to be age-related. (While we acknowledge various efforts to tackle these problems, for instance using self-supervised learning for the first, and learning with imbalanced data for the second, these are still challenging problems in machine learning). 
  Hence considering the general framework of machine learning-based casual inference in biological context \cite{b46} and the type of data at hand, we found it necessary to develop a new framework based on sample-analysis approach, for this type of data. Inspired by  the nature of our data, as well as popular 
  %famous 
  time series causal inference approaches \cite{b9,b10,b12}, here we formalize our hypothesis for causal inference.  

%\textcolor{red}{In one hand, complex patterns of variation in expression level of some of the age-associated  genes \cite{b11}, is a key observation for further pattern analysis. In particular,  variance  as  a  second  moment substantially reflects  the  complexity  and can be used to analyze the nature of complexity. The behaviour of this disturbance in terms of volatility in form of heteroskedasticity indicates underlying patterns with bi logical interpretation.}

On one hand, complex patterns of variation in expression levels of some of the age-associated  genes \cite{b11}, is a key observation that is important for further pattern analysis. In particular,  variance  as  a  second  moment substantially reflects  the  complexity  and can be used to analyze the nature of complexity. The behavior of this disturbance in terms of volatility in the form of heteroskedasticity could indicate important underlying patterns with biological significance.

On the other hand, 
%the  property  of conditional  volatility in the mechanism  of  human  aging. Specifically
%%from \cite{b11} we hypothesize that  gene expression data for those age-related genes \textbf{causing the aging (in contrast to those influenced by age)},
inspired by \cite{b11},  
we hypothesize that  gene expression data for those age-related genes \textbf{that cause aging (in contrast to those influenced by age)}, most probably follow a \textbf{consistent  volatility  pattern}  over the whole  human  lifespan.

%Taken all above mentioned together, 
Taken the above together, we hypothesize that, on the scale of human lifespan, consistent heteroskedastisity in age-related data can offer 
%offers great implication on causality, meaning that 
significant insights on probable causality. That is, we can single out genes with expression levels showing consistent hereroskedastisity over whole lifespan as those that are likely playing a causal role in aging. Later we extend the idea to explore non-causal age-related genes (i.e., associative, but not necessarily causal, genes).

%\section{Methodology}
% \vspace{-.7em}
%\label{sec:pagestyle}
%In this section we present the pipeline of the work as depicted in  Fig. \ref{Fig.001} and discuss each part separately.
Fig. \ref{Fig.001} shows the pipeline for our approach.  To efficiently perform the test of heteroskedasticity, first we model the vectors of gene expression data for each gene individually, then we extract the residual of the model for the needed test.
%of heteroskedasticity. 
ARIMA models are broadly known for their capability in capturing the linear behaviour of a variable \cite{b19}. ARCH test next will be performed on the residuals of the ARIMA model to detect the possible heteroskedasticity property of the time series.
% \vspace{-1.2em}
\subsection{ARIMA model}
% \vspace{-.7em}
Autoregressive integrated moving average (ARIMA) models are a set of statistical models to analyse  time series showing non-stationarity, as our data of gene expression presents this property. This class of models allows accurate capture of linear characteristics of the time series, and consequently provides  high quality residuals 
%in terms of non-linearity 
for further nonlinear testing or modeling, for instance, using the as ARCH test \cite{b22}.  

ARIMA models are generalized ARMA (autoregressive moving average) models, where an integrating part is added to generalize it to non-stationary time series modeling. The integrated (I) part of ARIMA is performed by a differencing step which aims at reducing non-stationarity \cite{b20,b19}. Let's say ${X_t}$ is the time series (${X_t}$ are real numbers and ${t}$ is integer index) corresponding to expression values of a certain gene spanning from age $<$ 1 year to age 94 years. Then ${ARIMA(p,d,q)}$ with ${p}$, ${q}$ and ${d}$, respectively, as the order of AR, MA and differencing part is given as:
% \vspace{-1.2em}
\begin{equation}
\label{eq:1}
(1-\sum_{i=1}^{p}a_iL^i)(1-L)^dX_t=(1+\sum_{i=1}^{q}b_iL^i) \epsilon_t
% \vspace{-1.5em}
\end{equation}
%\vspace{-0.2em}
where ${a_i}$ and ${b_i}$ represent the parameters of the autoregressive (AR) part of the model and the moving average (MA)  part, respectively, $L$ denotes the lag operator ($L\{O_t\}=O_t-O_{t-1}$)and ${\epsilon_t}$ the error term. A practical consideration is that the error terms ${\epsilon_t}$'s are generally presumed to be independent, and identically distributed (IID) variables which are sampled from a normal distribution with zero mean. To fit the ${ARIMA(p, d, q)}$ model with the best orders ${p}$, ${d}$ and ${q}$, the central idea is to find the most parsimonious ${ARIMA(p, d, q)}$, which practically is achieved  using  the corrected Akaike information criterion (${AIC_c}$). That is,  the best model introduces the minimum 
%numerical 
errors into the modeling and the standard deviation of the ARIMA model \cite{b21}. Table 1 shows the best model selections based on ${AIC_c}$.
% \vspace{-1.2em}
\begin{table}[h]
	\centering{
		\caption{Parameters of the ARIMA models for a sample modeled gene, as well as model ranking for all 550 age-related genes based on average ${AIC_c}$. 
		%Here $N (=143)$ is the number of expression level samples for a typical gene, for computing ${AIC_c}$. 
		Smaller ${AIC_c}$ values indicate better performance. 
		%, thus the best model, is associated with ${ARIMA(2, 1, 1)}$. 
		Our results on all 550 genes, show that, for most genes, the best models are ${ARIMA(2, 1, 1)}$, ${ARIMA(2, 1, 2)}$ and ${ARIMA(1, 1, 1)}$.}
		\label{table_1}
		\begin{tabular}{c|c|c}
			\hline
	
			{Model} &  {${AIC_c}$} & {Model Ranking} \\
			\hline
			\hline
			
			ARIMA(1,1,1)           & -7.3311e+03         &      \textbf{3}\\
			\textbf{ARIMA(2,1,1)}           & \textbf{-7.3334e+03} & \textbf{1}\\
			ARIMA(2,1,2)           & -7.3301e+03               & \textbf{2}\\
			ARIMA(1,1,2)           & -7.3030e+03               & \textbf{4}\\
			ARIMA(2,2,2)           & -7.3027e+03                & -\\
			ARIMA(2,2,1)           & -7.2914e+03               & -\\
			ARIMA(1,2,2)           & -7.2977e+03             & -\\
			ARIMA(1,2,1)           & -7.3116e+03             & \textbf{5}\\
			
			%$N_{acq}$                         & $160(8500)$         \\
%			\bottomrule[1.5pt]
		\end{tabular}}
	\end{table}
% 	\vspace{-2.3em}
	%\subsubsec
\subsection{Heteroskedasticity and ARCH Test}
% \vspace{-.7em}
To describe the variance of the error term, a statistical model, namely,  autoregressive conditional heteroskedasticity (ARCH) was developed as the function of the error term from the previous time period \cite{b22}. Its huge advantage over previous methods on modeling volatility which eventually brought a Noble prize for the inventor, is replacing assumptions of constant volatility with conditional volatility, meaning that past volatility influence future volatility. 
%This property of conditional volatility is advantageous here, as the literature on mechanism of human aging \cite{b11,b1} suggests that gene expression data for those age-related genes causing the aging (in contrast to those influenced by age), most probably follow a consistent volatility pattern over whole human lifespan in terms of future variation conditioned on the past variation. 
This property of conditional volatility is advantageous here. Given the literature on mechanism of human aging \cite{b11,b1}, we expect that gene expression data for those age-related genes that cause aging (in contrast to those influenced by age), would most probably follow a consistent volatility pattern over whole human lifespan in terms of future variation conditioned on the past variation. 

%Accordingly, 
To check for the time-varying property of conditional volatility in the form of heteroskedasticity, we use the ARCH test. 
The ARCH test is built based on the fact that if the residuals $r[t]$ are heteroskedastic, the squared residuals ($r^2[t]$) will be  autocorrelated \cite{b23}. There are two well-known tests for heteroskedasticity. The first one, known as Portmanteau Q test, examines whether the squares of residuals are a sequence of white noise. The second one, which we used in this work, is proposed by Engle \cite{b22}, and uses the Lagrange Multiplier test. Engle  proposed the Lagrange Multiplier test in terms of fitting a linear regression model for the squared residuals and assessing the significance of the fitted model. Engle’s ARCH test assesses the null hypothesis that a series of residuals $r[t]$ exhibits no ARCH effect, that is, conditional heteroskedasticity, where the alternative is that an ARCH($L$) model describes the time series. ARCH($L$) is represented by:
% \vspace{-.7em}
\begin{equation}
r_t^2 = a_0 + a_1r_{t-1}^2+...+a_Lr_{t-L}^2 +e_t,
\label{eq:2}
% \vspace{-1.5em}
\end{equation}
where there is at least one ${a_j \neq 0}$ for ${ j = 0,..,L}$. Moreover, the test statistic, the Lagrange multiplier statistic is ${TR^2}$, where ${T}$ is the sample size and ${R^2}$ is the coefficient of determination from fitting the ARCH($L$) model for a number of lags ${L}$ via regression. The null hypothesis is represented as:
% \vspace{-.8em}
\begin{equation}
H_0: a_0=a_1= ... =a_L=0 
\label{eq:4}
\end{equation}
Under the null hypothesis, the asymptotic distribution of the test statistic is chi-square with $L$ degrees of freedom
% \vspace{-1.2em}
\section{Experiment and Results}
% \vspace{-1.2em}

\subsection{Performing the Test on the Data}
% \vspace{-.7em}
The data consists of vectors of gene expression data with length 143, as our time series. As shown in Fig. \ref{Fig.001}, we perform the test at three scales (whole sequence, young, and old individuals) on age-related genes with causal relationship, other age-related genes and genes not associated with aging. Next, we put forward inference of causality. %Briefly noting that in biology literature of aging, there is a consensus on 
Following Belsky et al \cite{b24} where they showed that age 38 is a critical age, in this work, we use a simple approximation, and consider age 40 as a threshold age, meaning that age 40 is broadly considered as a discriminating age between young and old \cite{b24}. Accordingly, first we perform the ARCH test on the whole time series, then on the subjects that are 40 or under, and finally those above 40 years old. The hypothetical criterion derived from a combination of literature as well as well-known causal inference and inference of association is that, consistent heteroskedastisity substantially implies causality whereas non-consistent heteroskedastisity implies association. However, as shown in Fig. \ref{Fig.0} homoskedastisity (as opposed to heteroskedastisity) is  not interpreted either way, meaning that there might be causal or associative genes with homoskedastic property in their time span gene expression vectors. Here we 
%it is uber important to
emphasize that, as the nature of data is complex, \textbf{ we do not claim} that only one property would capture all signatures of causality/association for all causal/ associative genes.  

In this work, a given gene is said to have  "consistent heteroskedasticity" if the heteroskedasticity property is found in both young and old subjects. By this definition, it is possible that, for a certain gene, heteroskedasticity may appear over a lifetime (whole sequence) but still not consistent over young and old populations, since this may be due to dominance by one age group. Such a gene may be age-associated, but not a causal gene, based on our criteria. 

We describe the three considerations below:

\textbf{Whole sequence:} For each gene, we model the whole time series of 143 measurements of expression level over life span (1-, 94+ years old) using the ${ARIMA(p, d, q)}$ with best adjusted orders for each time series, and then perform the ARCH test on the residuals of this model. 

\textbf{Young Individuals:} For each gene we only model 70 measurements of the of expression level over young people (1-, 40 years old) and then perform the ARCH test on the residuals of the best ${ARIMA(p, d, q)}$ model. 

\textbf{Old Individuals:} Finally, we perform the test on the 73 measurements of the of expression level over old people (40+, 94+ years old) for each gene.

The whole process of testing is performed for four sets of genes: genes causing the aging process, genes influenced by aging, age-related genes for which the direction of association is unknown, and genes that are not known to be associated with aging.

% \subsection{Cointegration Test }
% \vspace{-.7em}
\subsection{Results}
\label{sec:Results}
% \vspace{-.7em}
%In this section, we present our results on both sets of genes, one with 69 genes known to influence ageing, and the other with 550 age-related genes (275+275). 

\begin{figure}[hbt]
	\centering{
		\includegraphics[width=3.6in]{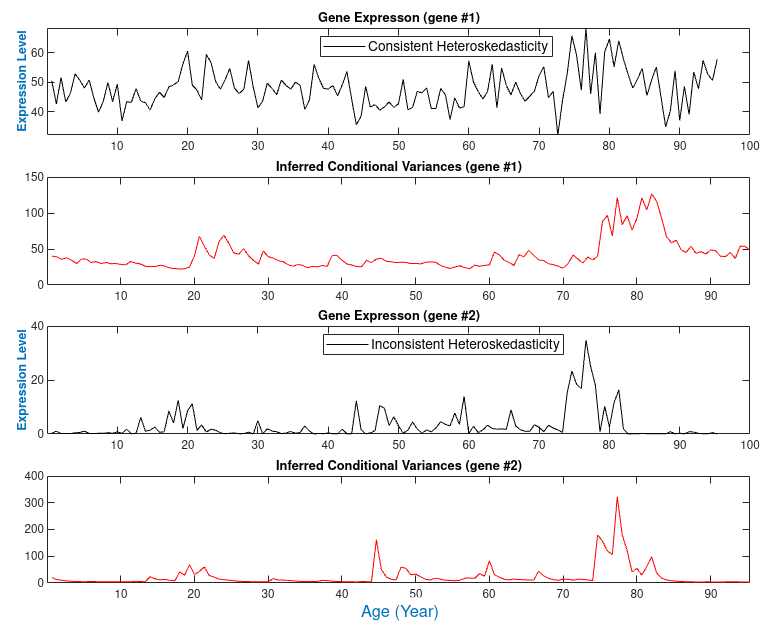}
		%\includegraphics[scale=1]{Fig1}
% 		\vspace{-1.em}

		\caption{Gene expression across age for two age-related genes. Top:  expression values for gene ANXA3 with causal relationship with aging showing consistent heteroskedasticity over the whole life span; Bottom: gene HLA-DRB5, an age-affected gene showing  inconsistent heteroskedasticity. 
		% \textcolor{red}{For this figure, everyone knows about the explosive growth of COVID-19 cases or deaths. For this paper, it will be better to show the growth of available COVID-19 sequences, and or the growth of COVID-19 variants.
		}
% 		\vspace{-1.2em}
		\label{Fig.2}}
%		\vspace{-1.6}
\end{figure}

Figure \ref{Fig.2} shows gene expression levels for two sample genes, one with consistent hetroskedasticity, and the other with inconsistent hetroskedasticity. In our model, consistency in  hetroskedasticity across young and old is used to detect causal age-related genes. 
Table \ref{table_Results_A} shows the results. This table shows the results of ARCH test as an indicator of the presence of heteroskedasticity, on expression values of different sets of genes over young individuals (${\leq}$ 40 years old), old individuals (over 40 years old), and whole life span (from less than 1 to 94 years). 

% \begin{figure}[t]
% 	\centering{
% 		\includegraphics[scale=.3]{AssoGeneExp.png}
% 		%\includegraphics[scale=1]{Fig1}
% 		\vspace{-2.em}

% 		\caption{Expression level across age for a typical gene
% 		% \textcolor{red}{For this figure, everyone knows about the explosive growth of COVID-19 cases or deaths. For this paper, it will be better to show the growth of available COVID-19 sequences, and or the growth of COVID-19 variants.
% 		\label{Fig.3}}}
% \end{figure}

% \vspace{-1.2}
%Table \ref{table_Results_A} shows the results. This table shows the results of ARCH test as an indicator of the presence of heteroskedasticity, on expression values of different sets of genes over young individuals (${\leq}$ 40 years old), old individuals (over 40 years old), and whole life span (from 0 to 94 years). 
%In the table, Non-significant and Significant genes are, respectively, age-related genes that we considered as affected genes by aging and  genes that cause aging based on the prior studies \cite{b11,b1}.
% \textcolor{red}{How do you define the 5000 age-unrelated genes?? Why do we say they are  age-unrelated?}

% \textcolor{blue}{How about the result on the remaining 27K-550-69 genes, which are not known to be associated with aging? Please include equivalent results for these in the table. These can represent potential novel age-related genes that is identified by the new method.}

Table \ref{table_Results_A} shows the results of our proposed approach using our datasets. 
As shown in Table \ref{table_Results_A}, \textbf{for a given gene}, presence of heteroskedasticity on both young and old subjects, is \textbf{at least one indicator of probable causal relationship with aging}. Clearly, the single indicator, the consistent heteroskedasticity proposed here \textbf{may not capture every casual gene}. The table shows that 12 potentially novel causal genes are identified (last row). 
The table also shows the results for age-associated genes identified using the heteroskedasticity test (see the last column). The results indicate that our proposed method using heteroskedasticity tests identified 547 potentially novel age-related genes, different from what is currently known in the literature. 

\begin{table}[h]
% \vspace{-1.2em}
	\centering{
		\caption{Results of heteroskedasticity test on different gene sets. Entries indicate \# of genes that passed the test. Y denotes Young, O: Old, L: Lifespan. Y\&O: intersection of Y and O, Union: union of Y,  O and L. Last two columns capture \# of causal genes and \# of age-associated genes, respectively. 
		%This table presents the results of ARCH test as an indicator of presence of heteroskedasticity, on expression values of different sets of genes over young individuals (less that 40 years old), old individuals (over 40 years old) and whole life span (less than 1 to 94 years old). Non-significant and significant genes are respectively age-related genes that we considered as affected genes by aging and  genes that cause aging based on the prior studies \cite{b11,b1}.
		}
		\label{table_Results_A}
		\footnotesize
		{
		\begin{tabular}{|c|c|c|c|c|c|}
			\hline
	
			{Gene Set} &  {Y} & {O} & L& \textbf{Y${\&}$O} & \textbf{Union}\\
			\hline
			\hline
			
			69 Causal genes           & 28   &     32      & 34 &28 & 34\\
			\hline
			275 Genes affected by aging            & 3 & 214 & 214 &0 & 217\\
			\hline
			275 Other age-related genes            & 82   &      92    &  97 & 57 & 134\\
			\hline
%			 5000 Non age-related genes           & 0      &     0   & 0 & 0\\
%			\hline
			 Remaining 26523 genes          & 31     &     501   & 513 & 12 & 547\\
			\hline
			
			%$N_{acq}$                         & $160(8500)$         \\
%			\bottomrule[1.5pt]
		\end{tabular}}
}
	\end{table}
	
Table \ref{table_Gene} shows more details on the 12 new genes identified by our approach as probable causal genes. The table provides supportive evidence for the 12 newly identified probable causal genes, based on the literature. Interestingly, our computational causal inference resulted in identifying new causal and associative age-related genes, some of which 
%that one can see some of them 
are already supported by wet-lab experimental evidence. Thus our work could help to substantially reduce the cost of wet lab experiments by guiding the experiments via some computational hints.    
% \vspace{-2.2em}
\section{ Discussion and further evaluation}

%\textcolor{red}{There are a few pints about the results and the dataset which are necessary to be clarified.}

Here, we briefly discuss the results and the dataset which may be helpful in placing the presented  work in context as well as evaluating the significance of the results.

\textbf{Dataset:} The datasets we used represent the only available dataset at this scale. Accordingly, in order to better evaluate the performance of our work, we have extended the experiment to genes without any label regarding existence of potential association with aging. Hence, the hypothesis testing was performed on all 27,142 genes. One point worth mentioning is that our method computationally fit the size of data, while using more complex models such as neural networks would require much more data to be trained.

\textbf{Association and causality:} Other causal inference work on gene expression data and aging are primarily 
wet lab experiments, hence we have no similar work to directly compare against. Developing a benchmark for further studies on computational causal inference in this problem domain is an important future work we plan to address.

Roughly speaking, in complex biological processes such as aging, there could be more than one indicator of causality. That is, one can not expect that in our problem setting, each of the potential indicators will predict or detect every causal gene in the genome. Rather, what we have presented is an approach to identify potentially causal genes as well as associative genes. The results showed that about half of the causal genes can be detected using this proposed computational approach. We see this as only the beginning, and we expect that other complimentary approaches will be needed for a more complete picture of the causal genes.

% \textcolor{blue}{Please provide more discussion if you find necessary, as we have one more page. Thanks.}
\begin{table}[h]
% \small
\vspace{-1.2em}
	\centering{
		\caption{The 12 identified probable causal genes. Prior studies suggest  potential connection with aging for most of the genes.}
		\label{table_Gene}
% 		\footnotesize
		{
		\begin{tabular}{|p{1.30cm}|p{6.9cm}|}
			\hline
	
			 \textbf{Gene} & \textbf{Comments/supportive evidence} \\
			\hline
			\hline
			 MFAP1 &  Evolution of the aging brain transcriptome \cite{b31}\\
			 \hline
			 CFAP45  &AMP binding activity, flagellated sperm motility \cite{b32} \\
			\hline
			 COMMD10 & Inhibits tumor progression (good for longevity) \cite{b33}   \\
			 \hline
			 ZNF558 & Age-related differential expression in breast cancer \cite{b34}   \\
			\hline
			TMPPE &  Linked to Gm1 Gangliosidosis disease; hydrolase activity\cite{b35}  \\
			\hline
			HTRA1 & Susceptibility to age-related macular degeneration type 7\cite{b36} \\
			\hline
			 STAG1 & Mutations linked to syndromic intellectual disability \cite{b37}   \\
			 \hline
			 MASCRNA & Related to gene Malat1  \cite{b38}, which is involved in cancer metastasis, cell cycle regulation \cite{b39}\\

			\hline
		 MARCH9 & Signaling vesicular transport 
		 %between  membrane compartments 
		 \cite{b40}  \\
		 \hline
		 CLDN22 & Maintaining cell polarity and signal transductions \cite{b41} \\
			\hline
			 PDE5A & 
			 Signal transduction; 
			 %by regulating the intracellular concentration of cyclic nucleotides. 
			 Muscle relaxation in the cardiovascular system \cite{b42}. Same family with an age-related gene \cite{b43}  \\
			 \hline
			 LUM & Linked to  gastric cancer tumorigenesis  \cite{b44}, collagen fibril organization, and circumferential growth \cite{b45}  \\
			\hline
			
			%$N_{acq}$                         & $160(8500)$         \\
%			\bottomrule[1.5pt]
		\end{tabular}}
}
	\end{table}
\textbf{Further evaluation:} 
% PDF estimation and KL divergence 
Undoubtedly, the best way to assess the results of such computational framework, is through wet-lab experiments and clinical studies. Hence we investigate the availability of possible hints in the literature and valid databases to verify our results. Table \ref{table_Gene} represents part of this investigation.\\
% \textcolor{red}{Regarding Table \ref{table_Gene}, 12 genes which are computationally identified as causal genes ...  }

\textbf{Significance of the results:}  As shown in Fig. \ref{Fig.0}, it is important to note the direction of the hypotheses. More generally, the problem of identifying causal/ associative age-related genes via data-driven computational methods, is limited to identifying causal/ associative age-related genes that have one property, while different causal/ associative age-related genes might manifest  other properties.
%computational property. 
For example, not all age-related genes might show variations in expression values, hence, even identifying a subset of causal/associative genes is 
%a great start with 
still an important contribution for this type of computational work. Further, no matter how complex and capable a method is, wet-lab experimental  verification is important to authenticate the results. Accordingly we investigated the experimental hints in the literature and elsewhere to assess and potentially authenticate our results resorting to the experimental studies. 

\textbf{Limitation of this work:} The results of this work and in general any work established based on analysis of genes expression variations is basically limited to identifying genes with variation in expression values, here  limited to causal/associative genes with variation in expression values. However, not all age-related genes might show variations in expression values. Therefore, for those genes with almost no variation in expression values, our proposed computational approaches may not be effective.  We mention that this will also depend on the specific experimental condition for which the gene expression data is captured. The approach proposed here can apply to other types of gene expression data, captured over age, and not only for the dermal fibroblast gene expression data used in this work.

\section{Conclusion}
% \vspace{-1.4em}
In this paper, we investigated potential causal and associative relationships between age-related genes and aging. Our central idea revolves around detecting heteroskedastic property in gene expression variation at different age ranges. Specifically, some genes that cause aging, as opposed to those influenced by aging would show the causality in terms of consistent heteroskedasticity, while other non-causal age-associated genes would express non-consistent heteroskedasticity in their expression values over the lifespan. That is, causal genes show heteroskedasticity in gene expression levels across both young and old. This is supported by the results of our experiments. Three main contributions of this work include developing a framework for causal and associative inference on age-related genes using gene expression data, identifying causal genes using consistent heteroskedasticity, and extending the framework to identify associative genes by presence of non-consistent heteroskedasticity. Supportive evidence from the literature on related wet-lab experiments provide an initial validation of some results from our proposed approach.

\section{Acknowledgements}
This work is supported in part by the US National Science Foundation (NSF), Award \#s: 1920920, 2125872.

\end{document}